\documentclass[aps,pra,twocolumn,superscriptaddress,showpacs,nofootinbib]{revtex4-1}
\usepackage{graphicx,lipsum}
\usepackage{epstopdf}
\usepackage{subfigure}
\usepackage{amsmath,color}
\def\gappeq{\mathrel{ \rlap{\raise.5ex\hbox{$>$}}
                      {\lower.5ex\hbox{$\sim$}} } }
\def\lappeq{\mathrel{ \rlap{\raise.5ex\hbox{$<$}}
                      {\lower.5ex\hbox{$\sim$}} } }
\usepackage{bm}
\usepackage{color}

\newcommand\di{\text{d}}

\begin{document}

\title{Exploring the Stability and Dynamics of Dipolar Matter-Wave Dark Solitons}

\author{M. J. Edmonds}
\affiliation{Joint Quantum Centre Durham--Newcastle, School of Mathematics and Statistics, Newcastle University, Newcastle upon Tyne, NE1 7RU, United Kingdom}
\author{T. Bland} 
\affiliation{Joint Quantum Centre Durham--Newcastle, School of Mathematics and Statistics, Newcastle University, Newcastle upon Tyne, NE1 7RU, United Kingdom}
\author{D. H. J. O'Dell}
\affiliation{Department of Physics and Astronomy, McMaster University, Hamilton, Ontario, L8S 4M1, Canada}
\author{N. G. Parker}
\affiliation{Joint Quantum Centre Durham--Newcastle, School of Mathematics and Statistics, Newcastle University, Newcastle upon Tyne, NE1 7RU, United Kingdom}


\begin{abstract}
We study the stability, form and interaction of single and multiple dark solitons in quasi-one-dimensional dipolar Bose-Einstein condensates. The  solitons are found numerically as stationary solutions in the moving frame of a non-local Gross Pitaevskii equation, and characterized as a function of the key experimental parameters, namely the ratio of the dipolar atomic interactions to the van der Waals interactions, the polarization angle and the condensate width. The solutions and their integrals of motion are strongly affected by the phonon and roton instabilities of the system. Dipolar matter-wave dark solitons propagate without dispersion, and collide elastically away from these instabilities, with the dipolar interactions contributing an additional repulsion or attraction to the soliton-soliton interaction. However, close to the instabilities, the collisions are weakly dissipative.

\end{abstract}

\maketitle

\section{Introduction}

Solitary waves, or solitons, are excitations of nonlinear systems that possess both wave-like and particle-like qualities. They obey wave equations and yet do not disperse, maintaining their shape and speed by balancing dispersion with nonlinearity.  
Solitons appear across a wide range of physical systems that include water, light, plasmas and liquid crystals \cite{dauxois_2006}, and have been touted as playing a fundamental role in the fabric of our universe \cite{manton_2007}. A more recent addition to this list is the atomic Bose-Einstein condensate (BEC) formed in an ultracold gases which are described by a nonlinear Schr\"{o}dinger equation known as the Gross-Pitaevskii equation (GPE). Experimental demonstrations of solitons in BECs include both the generation of dark solitons \cite{Burger1999,Denschlag2000,Dutton2001,Engels2007,Jo2007,Becker2008,Chang2008,Stellmer2008,Weller2008}, and bright solitons \cite{strecker_2002,khaykovich_2002,eiermann_2004,cornish_2006, marchant_2013,mcdonald_2014}. The interaction between the atoms in these experiments was short range and isotropic (predominantly of the van der Waals type) giving a {\it local} cubic nonlinearity in the GPE, with dark and bright solitons supported for repulsive and attractive nonlinearity, respectively.  In binary condensates, dark-bright soliton complexes have also been probed experimentally \cite{Becker2008,hamner_2011}. The experiments confirm that solitons in BECs and in classical systems such as water or light are for many purposes the same phenomenon and share the same three particle-like defining properties, namely: permanent form, localization within a region, and emergence from collisions with other solitons unaltered, except for a phase shift \cite{drazin_1989}. It is important to bear in mind, however, that in BECs the soliton relies on quantum mechanical coherence across the sample, and is at heart a probability wave \cite{bloch_2008}.

Ultracold Bose gases provide an appealing system in which to explore soliton physics because of the almost complete absence of dissipation (due to the superfluid nature of the gas) and the high degree of experimental control that can be exerted over the atoms and their interactions using lasers as well as magnetic and electric fields. This, for example, has led to proposals to access exotic solitons such as in spin-orbit coupled condensates \cite{achilleos_2013a,achilleos_2013b,xu_2013}, and chiral solitons in `interacting' gauge theories \cite{edmonds_2013}, and non-local solitons in dipolar condensates \cite{cuevas_2009,baizakov_2015,bland_2015,pawlowski_2015} which are the subject of this paper.  Furthermore, matter-wave solitons have been proposed for  applications in precision interferometry \cite{strecker_2002,negretti_2004,mcdonald_2014,helm_2015} and surface interrogation \cite{cornish_2008}.
The experimental achievement of condensation of bosonic elements possessing large magnetic dipole moments --- $^{52}$Cr \cite{griesmaier_2005,beaufils_2008}, $^{164}$Dy \cite{lu_2011,tang_2015} and $^{168}$Er \cite{aikawa_2012} --- has opened yet another chapter in BEC physics. Dipoles introduce long-range anisotropic interactions falling off as $1/r^{3}$, in contrast to the usual short-range isotropic interactions, and hence give rise to an additional {\it non-local} nonlinearity \cite{lahaye_2009}.     
This has striking consequences, as observed experimentally in the form of magnetostriction of the condensate \cite{stuhler_2005} and shape-dependent stability \cite{koch_2008}, anisotropic collapse and explosion \cite{lahaye_2008}, and droplet formation analogous to the Rosensweig instability in classical ferrofluids \cite{kadau_2015,barbut_2016}. This modulational instability is a direct result of the roton dip the dipolar interactions introduce into the excitation spectrum \cite{santos_2003,giovanazzi_2004}.

\begin{figure}[b]
\includegraphics[scale=0.5]{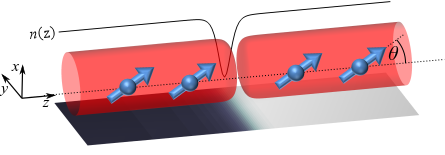}
\caption{(Color online) We consider an elongated Bose-Einstein condensate of atoms (shown as a density isosurface) with dipole moments (shown as arrows) which are co-polarized in a common direction. For suitable interaction parameters, a dark soliton can be supported, characterized by a 1D density notch and a non-trivial 1D phase slip.}
\label{fig:cigar}
\end{figure} 

Another prominent physical system that supports dark solitons with nonlocal interactions are nonlinear optical media, where the interaction of the electric field of light with the material gives rise to a defocussing local nonlinearity \cite{kivshar_1998}, and a nonlocal nonlinearity can arise due to thermal conduction.
This is typically modeled via a response function which decays exponentially with distance \cite{nikolov_2003,kong_2010}. A strong mathematical analogy can be drawn between these optical systems and atomic gas condensates, since both are studied with a similar underlying model, at least in the purely local case \cite{proukakis_2004}. Studies of the optical systems with nonlocal nonlinearity have focused on their stability \cite{krolikowski_2004}, and the arising interaction forces between dark solitons \cite{nikolov_2004,chen_2014}. This has in turn lead to the observation \cite{dreischuh_2006} of both repulsion and attraction between dark solitons, with the latter supporting bound states in the optical context.  The generation of dark solitons from shocks in these systems has also been experimentally studied \cite{conti}.  

The local cubic defocussing Schr\"odinger equation represents a solvable model within the framework of the inverse scattering method \cite{kivshar_1989,malomed_2005}. The inclusion of a cubic non-local potential to this model greatly complicates its analytical treatment within this method, there currently being no-known exact solutions. Nevertheless, approximate methods including series approximations for the non-local potential \cite{kong_2010a} and variational calculations \cite{pu_2012} have been successfully employed within this context to elucidate the physics of these models, with the latter capturing the existence of dark soliton bound-states.

A series of theoretical investigations have indicated that dipolar interactions in BECs also considerably enrich the properties of solitons. In quasi-one-dimensional geometries, bright \cite{cuevas_2009} and dark solitons \cite{pawlowski_2015,bland_2015} can be supported when the net (van der Waals and dipolar) interactions are attractive and repulsive, respectively.  Dark-in-bright and bright-in-dark dipolar solitons have also been predicted \cite{adhikari_2014}.  Yet perhaps the most interesting facet of solitons in general are their interactions and collisions \cite{stegeman_1999}, and in dipolar condensates the solitons themselves, considered as individual particle-like entities, inherit non-local soliton-soliton interactions in addition to the usual short-range soliton-soliton interaction  \cite{cuevas_2009,bland_2015}.  The play off between these two contributions can lead to the formation of unconventional bound states of bright \cite{baizakov_2015} and dark solitons \cite{pawlowski_2015,bland_2015}. Dipole-dipole interactions are also predicted to support two-dimensional bright solitons in quasi-2D geometries \cite{pedri_2005,tikhonenkov_2008,raghunandan_2015}, and suppress the well-known transverse ``snaking'' instability of dark solitons in three-dimensional geometries \cite{nath_2008}.




In a recent work \cite{bland_2015}, we predicted the existence of dark soliton solutions in a homogeneous quasi-one-dimensional dipolar condensate (shown schematically in Fig. 1), and studied the non-local interaction between two such solitons.  Here we expand upon this topic by presenting a comprehensive analysis of the family of dark soliton solutions and their interactions, across the main system parameters, namely the angle of polarization, the relative strength of the dipolar interactions, the soliton speed and the width of the quasi-1D system.  In order to understand the regimes of soliton stability, we establish the stability properties of the soliton-free ground state of the system; this allows us to relate the soliton stability to the phonon and roton instabilities of the system.  We examine the family of single soliton solutions, including their integrals of motion, and finally explore the soliton-soliton collisions through simulations of the dipolar GPE.

The main body of the paper is organized into four sections. In Sec.~\ref{sec:mf1d} we derive the mean-field equation of motion for the dipolar condensate. Following this in Sec.~\ref{sec:inst} we analyze the homogeneous system, obtaining analytical expressions for the position of the phonon and roton instabilities. Section \ref{sec:dds} describes single dipolar dark soliton properties and solutions across the full parameter space of the problem, while Sec.~\ref{sec:dsint} explores their collision dynamics. Our findings are summarized in the conclusion, Sec. \ref{sec:conc}. The body of the paper is supported by a technical appendix explaining the numerical method used to obtain the dark soliton solutions.


\section{Mean-Field Model of the Dipolar Condensate \label{sec:mf1d}}
We consider a gaseous BEC of ultracold weakly-interacting atoms with mass $m$ and permanent magnetic dipole moment $d$. Within the mean-field Gross-Pitaevskii theory, the atom-atom interaction in the low energy limit is described by the pseudo-potential
\begin{equation}\label{eqn:u3d}
U({\bf r}-{\bf r}')=g\delta({\bf r}-{\bf r}')+U_{\text{dd}}({\bf r}-{\bf r}'). 
\end{equation}
The first term arises from the short range isotropic van der Waals-type interactions, where $g=4\pi\hbar^2a_s/m$ and $a_s$ defines the $s$-wave scattering length. The long-range and anisotropic interaction appears as the bare dipole-dipole interaction between point dipoles \cite{santos_2000,yi_2001}
\begin{equation}
U_{\text{dd}}(r)=\frac{C_{\text{dd}}}{4\pi}\hat{e}_{j}\hat{e}_{k}\frac{(\delta_{jk}-3\hat{r}_j\hat{r}_k)}{r^3},
\label{eqn:dd3d}
\end{equation} 
where $C_{\text{dd}}=4\pi d^{2}$ characterizes the strength of the dipole-dipole interaction and $\hat{e}_j$ is the unit vector in the coordinate direction $\hat{r}_j$. Equation \eqref{eqn:dd3d} can also be written as
\begin{equation}\label{eqn:ddalt}
U_{\text{dd}}({\bf r}-{\bf r}')=\frac{C_{\text{dd}}}{4\pi}\frac{1-3\cos^2\theta}{|{\bf r}-{\bf r}'|^3},
\end{equation}
where the angle $\theta$ is defined between the vector joining the dipoles and the polarization direction (see Fig.~\ref{fig:cigar}). 
At the so called magic angle $\theta_m\simeq54^{\circ}$, the dipole-dipole interaction reduces to zero. Assuming $C_{\text{dd}}>0$, then for $\theta<\theta_m$ (dipoles orientated predominantly head-to-tail) the dipolar interaction is attractive. For $\theta>\theta_m$  (dipoles dominantly side-by-side) the interaction is repulsive. It is also possible to consider a regime of ``anti-dipoles'', $C_{\text{dd}}<0$, as proposed in Ref.~\cite{giovanazzi_2002} by rapidly rotating the dipoles, for which the attractive and repulsive regimes are reversed.  It is convenient to specify the relative strength of the dipole-dipole interaction to the van der Waals interaction via the parameter $\varepsilon_{\text{dd}}=C_{\text{dd}}/3g$.  By means of Feshbach resonances to tune $g$ \cite{lahaye_2007}, as well as the above mentioned method of generating $C_{\rm dd}<0$, it is experimentally possible to access systems over the full range $-\infty <\varepsilon_{\text{dd}}< \infty$, with negative or positive $g$ or $C_{\rm dd}$. 

The quantum state of the dipolar condensate is described by its  mean-field wavefunction $\Psi({\bf r},t)$; the condensate density distribution follows as $n({\bf r})=|\Psi({\bf r},t)|^2$. In the limit of zero temperature, the wavefunction obeys the dipolar GPE \cite{lahaye_2009}
\begin{equation}
i\hbar\frac{\partial\Psi}{\partial t}=\bigg[-\frac{\hbar^2}{2m}\nabla^2+V({\bf r})+g|\Psi({\bf r},t)|^2+\Phi({\bf r},t)\bigg]\Psi.
\label{eqn:dgpe}
\end{equation}
The external potential $V({\bf r})$ which confines the cloud can in general take many forms, but we assume it to be a harmonic waveguide given by $V({\bf r})=m\omega_{\bot}^{2}r_{\perp}^{2}/2$, where $r_{\perp}=\sqrt{x^2+y^2}$ defines the radial coordinate, and the transverse trapping frequency is $\omega_{\perp}$.  We neglect any axial confinement since the exact soliton solutions we seek exist only for axially uniform systems.  It is worth noting that such axially-uniform waveguides are accessible experimentally \cite{strecker_2002}.
Meanwhile, $\Phi_{\rm dd}({\bf r},t)$ is the non-local mean-field potential generated by the dipoles 
\begin{equation}
\Phi({\bf r},t)=\int \di{\bf r}'U_{\text{dd}}({\bf r}-{\bf r}')n({\bf r}',t).
\end{equation}

In this work we consider a quasi-one-dimensional dipolar condensate \cite{parker_2008}, under which the 3D GPE can be reduced to an effective 1D equation. The transverse harmonic trapping is sufficiently tight ($\hbar\omega_{\perp}\gg\mu$, where $\mu$ is the chemical potential of the three-dimensional system) that no transverse degrees of freedom are excited. The wavefunction is taken to follow the ansatz $\Psi({\bf r},t)=\psi_{\perp}(r_{\perp})\psi(z,t)$, where $\psi_{\perp}(r_{\perp})=(l_{\perp}\sqrt{\pi})^{-1}\exp(-r_{\perp}^{2}/2l_{\perp}^{2})$, and $l_{\perp}=\sqrt{\hbar/m\omega_{\perp}}$ defines the transverse harmonic length scale. Such a state constitutes a single mode approximation (SMA) for the axial dynamics of the condensate.
Inserting this ansatz into the 3D GPE \eqref{eqn:dgpe} and integrating out the transverse ($x$-$y$) dimensions leads to the effective 1D dipolar GPE
\begin{equation}
i\hbar\frac{\partial\psi}{\partial t}=\bigg[-\frac{\hbar^2}{2m}\frac{\partial^2}{\partial z^2}+\frac{g}{2\pi l_{\perp}^{2}}|\psi|^2+\Phi_{\text{1D}}(z,t)\bigg]\psi.
\label{eqn:gpe1d}
\end{equation}
The effective 1D dipolar mean-field potential is
\begin{equation}
\Phi_{\text{1D}}(z,t)=\int dz' U_{\text{1D}}(z-z')|\psi(z',t)|^2,
\end{equation}
where the effective dipolar pseudo-potential is $U_{\text{1D}}(z)=U_{0}\tilde{U}_{\text{1D}}(|z|/l_{\perp})$ with \cite{deuretzbacher_2010,sinha_2007}
\begin{equation}
\tilde{U}_{\text{1D}}(u){=}\bigg[2u{-}\sqrt{2\pi}(1{+}u^2)e^{u^2/2}\text{erfc}\bigg(\frac{u}{\sqrt{2}}\bigg)\bigg]{+}\frac{8}{3}\delta(u).
\label{eqn:pp1d}
\end{equation}
The term in square brackets in Eq. \eqref{eqn:pp1d} gives a non-local contribution to the mean-field interactions, while the second contact-like term describes a short-ranged local contribution to the mean-field interactions. The strength and orientation of the dipole-dipole interaction is captured by $U_{0}=C_{\text{dd}}(1+3\cos2\theta)/32\pi l_{\perp}^{3}$.  It is also useful to know the energy of the condensate, which is given by the integral
\begin{equation}
E=\int^{\infty}_{-\infty} \di z \left[\frac{\hbar^2}{2m}\bigg|\frac{\partial\psi}{\partial z}\bigg|^2+\frac{g}{4\pi l_\perp^2}|\psi|^4+\frac12\Phi_{\text{dd}}|\psi|^2\right].
\label{eqn:en}
\end{equation}
The three terms here represent kinetic energy, interaction energy due to vdWs interactions and interaction energy due to dipolar interactions.


From now on our analysis of dipolar dark solitons will be based on the effective 1D dipolar GPE \eqref{eqn:gpe1d}.  
Our results will be presented in terms of the natural quantities of the homogeneous (soliton-free) condensate.  Taking $n_0$ as the uniform density, the chemical potential (the energy eigenvalue of Eq. \eqref{eqn:gpe1d}) follows as $\mu_0=n_0g/(2\pi l_{\perp}^{2})+\Phi_0$.  Here the first term represents the van der Waals contribution, and $\Phi_0=-C_{\text{dd}}n_{0}[1+3\cos2\theta]/24\pi l_{\perp}^{2}$ is the dipolar contribution. 
The natural units of length and speed are the dipolar healing length, $\xi=\hbar/\sqrt{m\mu_0}$, and the speed of sound, $c=\sqrt{\mu_0/m}$. A time unit follows as $\tau=\xi/c$.  To parameterize the cross-over from three to one dimension we define $\sigma=l_{\perp}/\xi$; the quasi-one-dimensional limit is valid for $\sigma\lappeq 1$ \cite{gorlitz_2001}.

\section{Stability of the Homogeneous System\label{sec:inst}}
Dark solitons are excitations of a background condensate, and so the stability of these states is heavily influenced by that of the background condensate.  Here we perform an analysis of the stability of the homogeneous quasi-1D dipolar condensate.  This is a generalization of the results of Refs. \cite{giovanazzi_2004, sinha_2007} which only considered dipoles polarized along the long axis.  Although strictly speaking the dark soliton state will possess a different excitation spectrum to that calculated in this section, we will see that this simple analysis agrees remarkably well with the position of the instabilities of the dark soliton solutions to Eq.~\eqref{eqn:gpe1d}.  


\begin{figure}[t]
\includegraphics[scale=0.45]{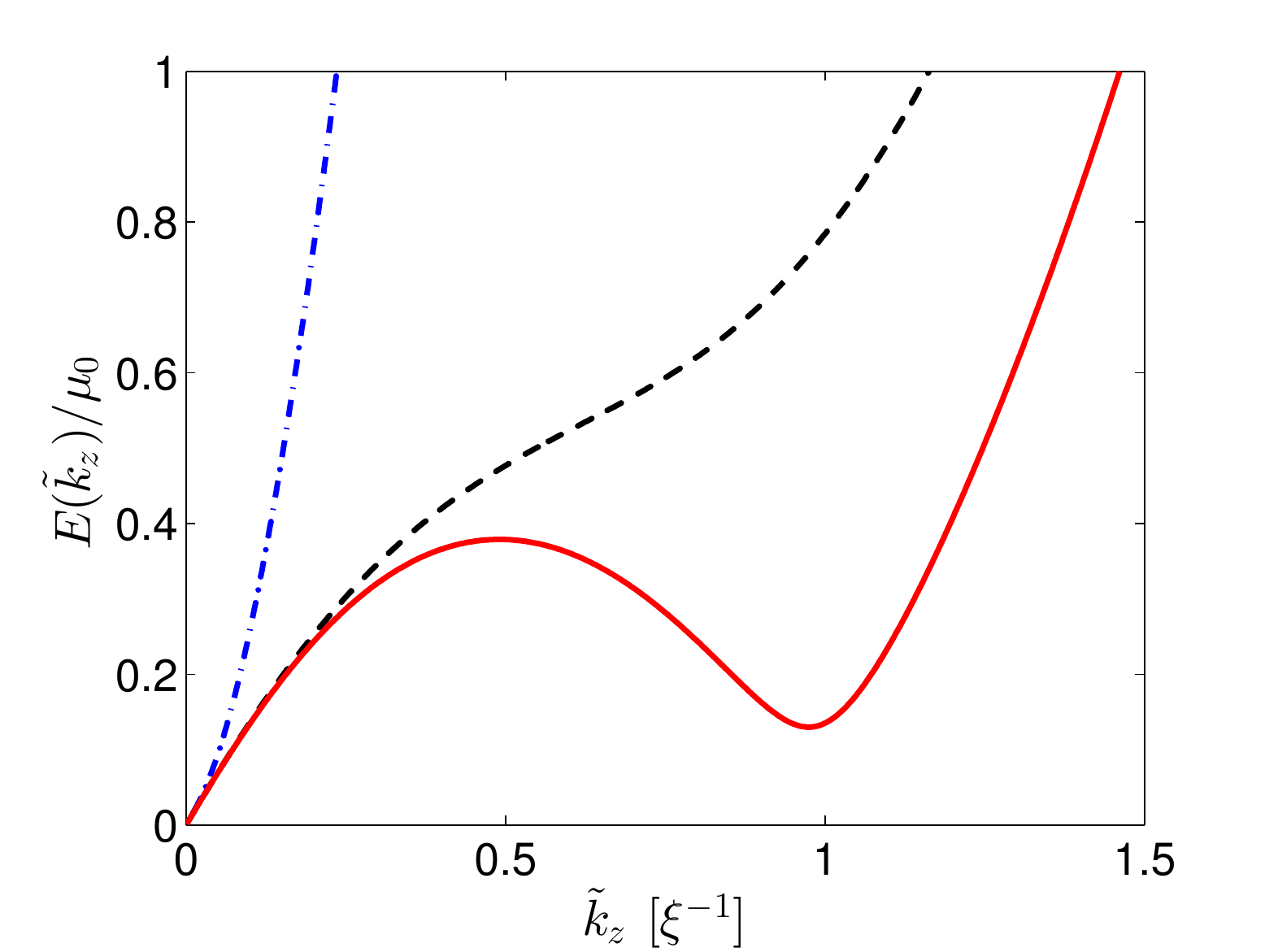}
\caption{(Color online) The Bogoluibov de gennes spectrum, Eq.~(\ref{eqn:bdg}), plotted for three illustrative values of $\varepsilon_{\text{dd}}$: $0.95,-2,-5$ (dot-dashed blue, dashed black, solid red).  These show, respectively, the conventional phonon/free-particle spectrum, the flattening off of the dispersion relation at intermediate $k$, and the emergence of a roton minimum.  The condensate has arbitrary width $\sigma=1$.}
\label{fig:bdg}
\end{figure}
The stability of the condensate can be deduced from the fate of small perturbations whose energies are given by the Bogoliubov excitation spectrum. For a homogeneous system, the Bogoliubov spectrum depends on the momentum (Fourier) space version of Eq. \eqref{eqn:pp1d}. However, rather than directly transforming Eq. \eqref{eqn:pp1d}, we can proceed more easily by  returning to the original 3D dipolar interaction given in Eq. \eqref{eqn:dd3d} and transform it into Fourier space by using the useful identity \cite{craig_1999}    
\begin{equation}
\frac{1}{4\pi r^{3}}\bigg(\delta_{jk}-3\hat{r}_{j}\hat{r}_{k}\bigg)=\frac{2}{3}\delta_{jk}\delta({\bf r})-\delta^{\perp}_{jk}({\bf r}),
\label{eqn:ddrel}
\end{equation}
where the quantity $\delta^{\perp}_{jk}({\bf r})$ is the transverse part of the delta function, defined as
\begin{equation}
\delta_{jk}^{\perp}({\bf r})=\int \frac{\di^{3}{\bf k}}{(2\pi)^{3}}e^{i{\bf k}\cdot{\bf r}}(\delta_{jk}-\hat{k}_{j}\hat{k}_{k}) .
\label{eqn:delta}
\end{equation} 
The quantity $\hat{k}_{j}$ appearing in Eq.~(\ref{eqn:delta}) is the $j$th component of the unit vector $\hat{\mathbf{k}}$ in Fourier space. Using these results together with the Fourier representation of the delta function, the Fourier transform of Eq. \eqref{eqn:dd3d} can then be directly found as
\begin{equation}
U_{\mathrm{dd}}(\mathbf{k})=\frac{C_{\mathrm{dd}}}{3} \hat{e}_{j}\hat{e}_{k}\left(3\hat{k}_{j}\hat{k}_{k}-\delta_{jk}\right).
\end{equation}
The dimensional reduction to 1D can be performed directly in Fourier space in an analogous way to the real-space reduction, again by assuming a harmonic ground state in the transverse directions. This yields the momentum space equivalent of Eq. \eqref{eqn:pp1d} \cite{giovanazzi_2004} 
\begin{equation}
\frac{U_{\text{1D}}(k_z)}{l_{\perp}}{=}4U_{0}\bigg[\frac{k_{z}^{2}l_{\perp}^{2}}{2}e^{k_{z}^{2}l_{\perp}^{2}/2}E_{1}\bigg(\frac{k_{z}^{2}l_{\perp}^{2}}{2}\bigg){-}1\bigg]{+}\frac{8}{3}U_{0},
\label{eqn:pp1dk}
\end{equation}
where $\hbar k_z$ is the momentum associated with the axial $z$-direction, while $E_{1}(x)=\int^{\infty}_{x}dt\ t^{-1} e^{-t}$ is the exponential integral. 
The total one-dimensional pseudo-potential, including van der Waals and dipolar contributions, is then  $U_{\text{tot}}(k_z)=g/(2\pi l_{\perp}^{2})+U_{\text{1D}}(k_z)$. Both Eq.~\eqref{eqn:pp1d} and \eqref{eqn:pp1dk} can be split into a non-local and local contribution; the former gives a contribution to the total contact interactions while the latter forms the important long-ranged part of the dipolar interaction. We note that in general the scattering length, and hence $g$, can be modified both by the dipolar interactions and confinement induced resonances (due to tight 1D trapping) and so may not take on their 3D dipole-free values \cite{sinha_2007}. Nevertheless, the general form of this pseudo-potential is expected to hold.

\begin{figure}[b]
\includegraphics[scale=0.55]{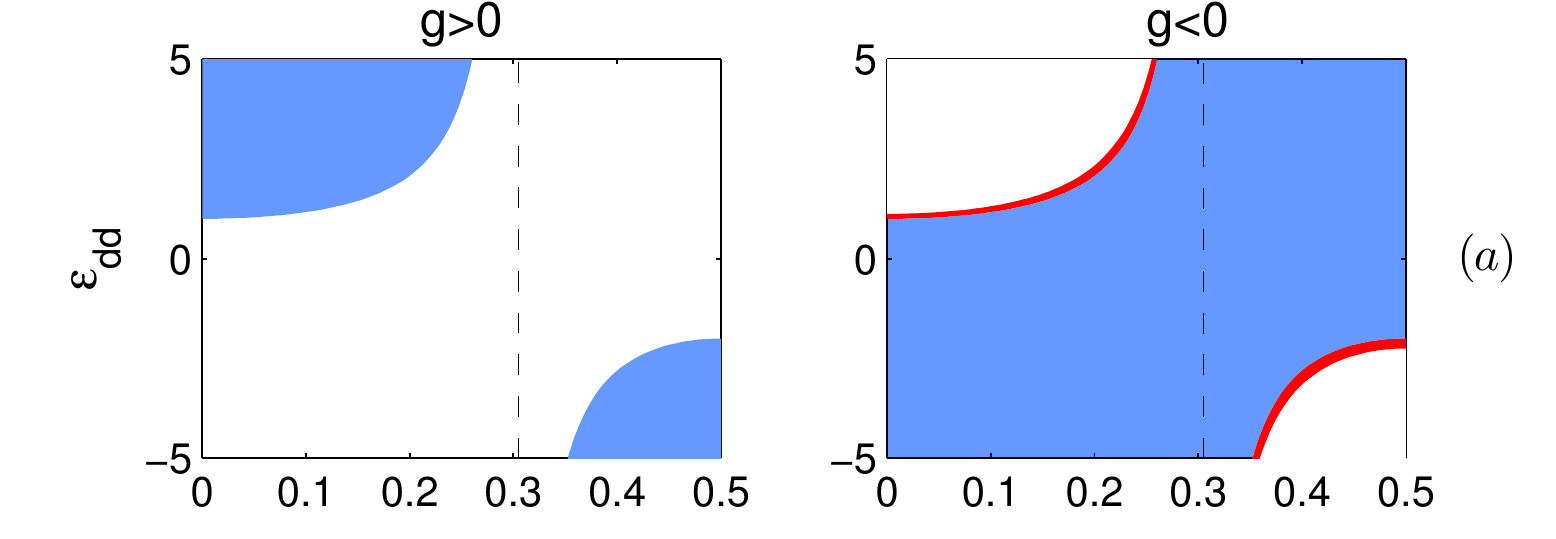}
\includegraphics[scale=0.55]{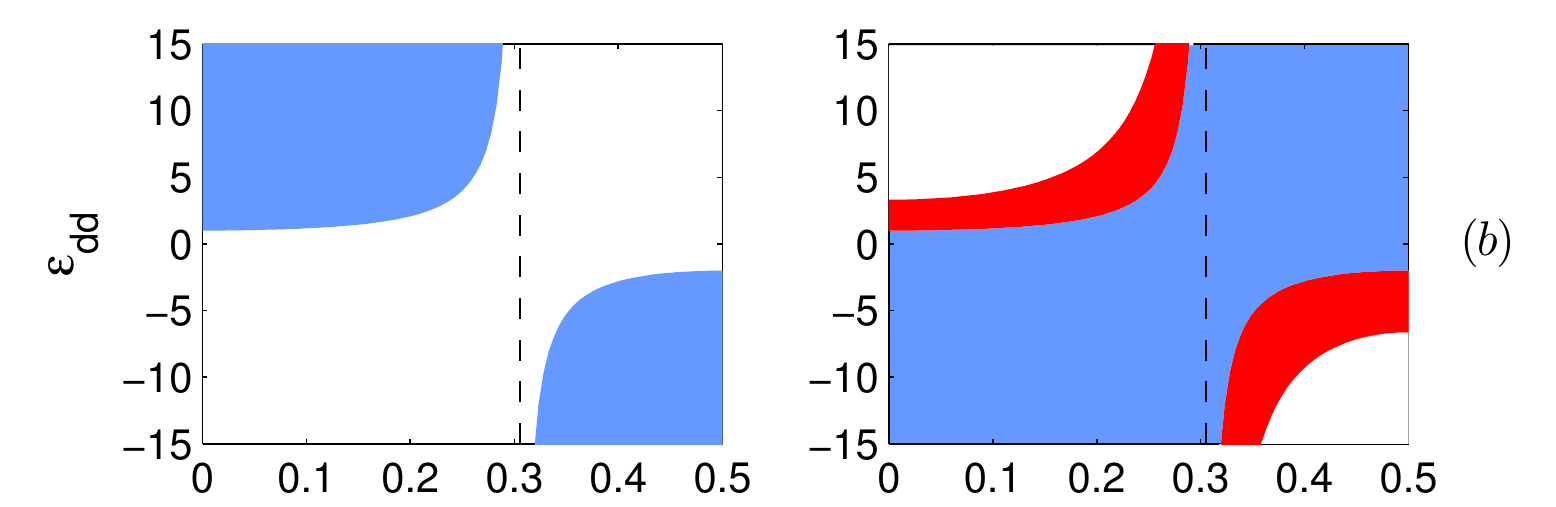}
\includegraphics[scale=0.55]{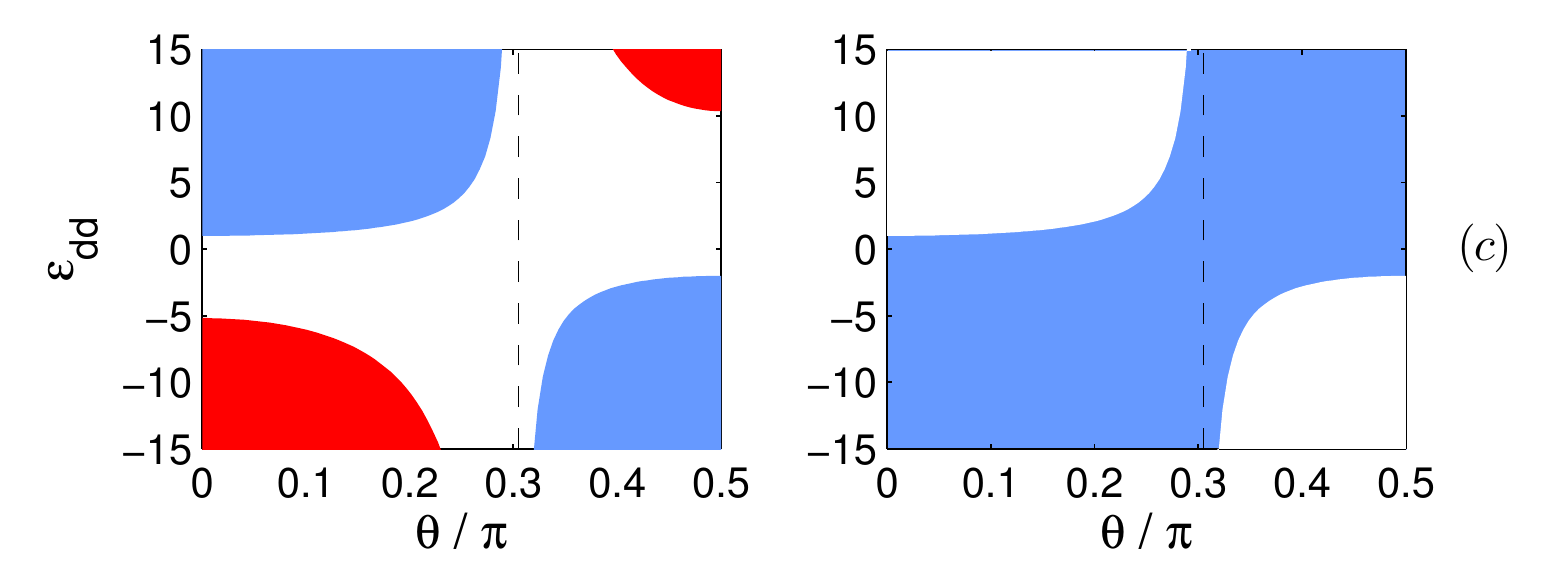}
\caption{(Color online) Stability diagrams corresponding to the homogeneous ground state of Eq. \eqref{eqn:gpe1d} in the $(\theta,\varepsilon_{\text{dd}})$ plane for $g>0$ (left column) and $g<0$ (right column). Rows (a), (b) and (c) correspond to $\sigma=0.1,0.5$ and $1.0$ respectively. The blue (red) bands represent regions of phonon (roton) instability, while white regions are stable. The magic angle, $\theta_m\simeq54^{\circ}$ is indicated in dashed black. Note that dipolar interactions can act to either destabilize or stabilize the BEC. The latter situation occurs when $g<0$ but the net dipolar interactions are repulsive.}
\label{fig:evt}
\end{figure}

With the above results in hand, the Bogoliubov excitation spectrum for perturbations of momentum $\hbar k_{z}$ from the background can be written as 
\begin{equation}
E_{k}^{2}=\epsilon_{k}^{2}+2n_{0}\epsilon_{k}\bigg[4l_{\perp}U_{0}\mathcal{V}_{\text{1D}}\bigg(\frac{k_{z}^{2}l_{\perp}^{2}}{2}\bigg)+\frac{g}{2\pi l_{\perp}^{2}}\bigg],
\label{eqn:bdg}
\end{equation} 
where $\epsilon_k={\hbar^2k_{z}^{2}/2m}$ defines the free particle energy and $\mathcal{V}_{\text{1D}}(q)=q\exp(q)E_{1}(q)-1/3$. The excitations associated with Eq.~(\ref{eqn:bdg}) are running waves of the form
\begin{equation}
\delta\psi(z,t){=}u(k_z)e^{-i(k_{z}z{-}E_{k}t/\hbar)}{+}v^{*}(k_z)e^{i(k_{z}z{-}E_{k}^{*}t/\hbar)}
\end{equation}
that constitute small amplitude fluctuations about the stationary condensate.

\begin{figure*}[t]
\includegraphics[scale=0.37]{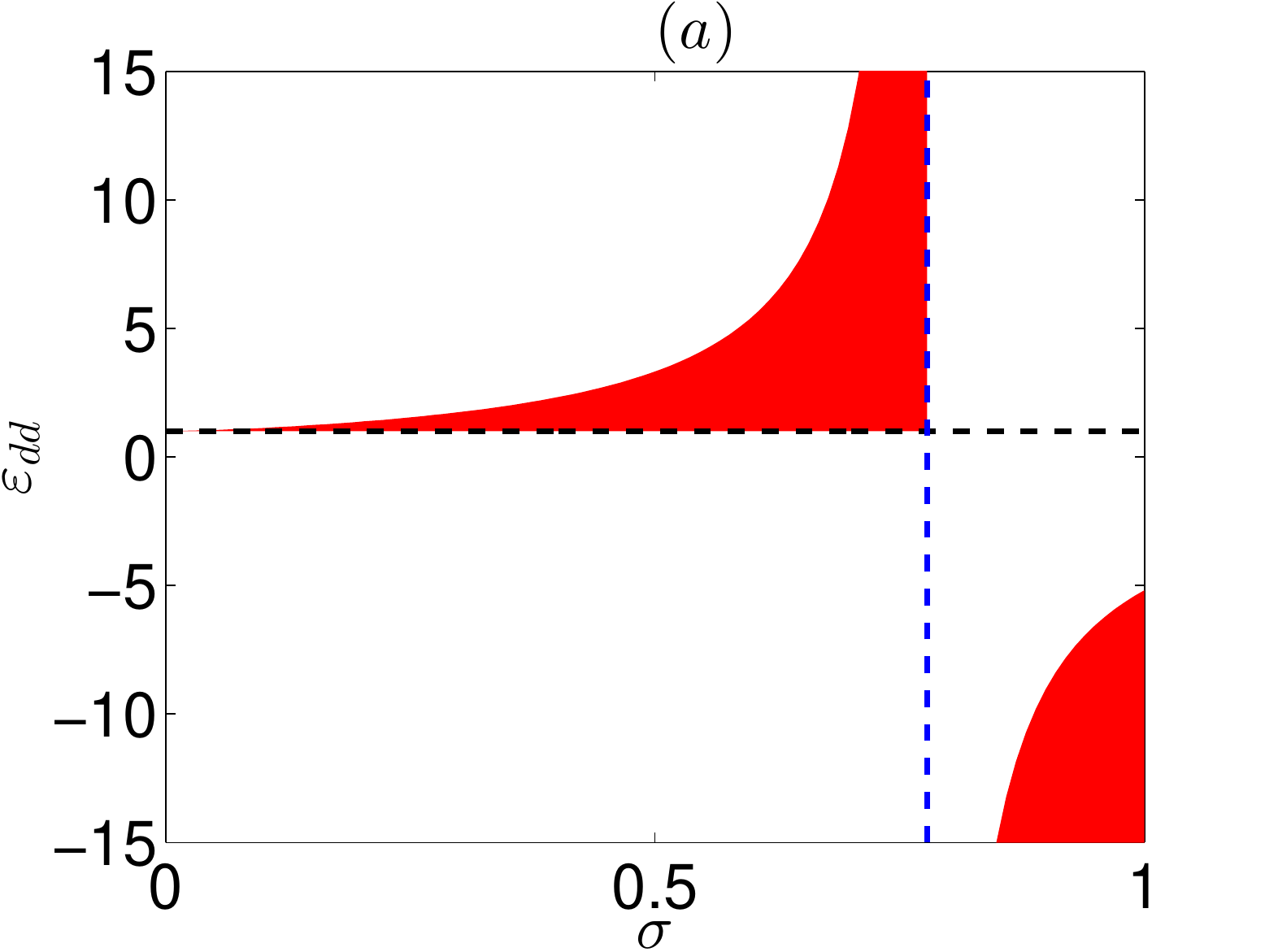}
\includegraphics[scale=0.37]{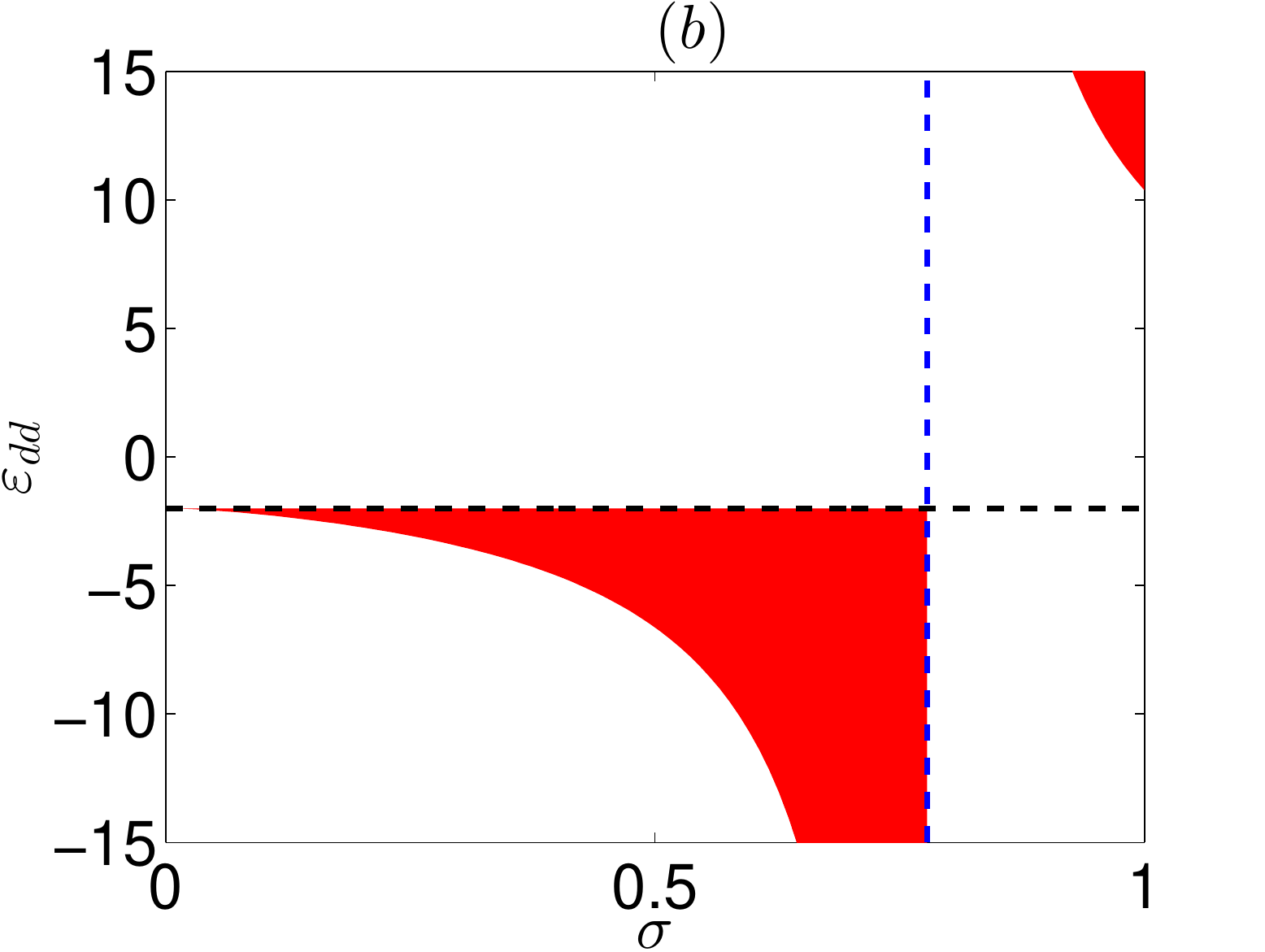}
\includegraphics[scale=0.37]{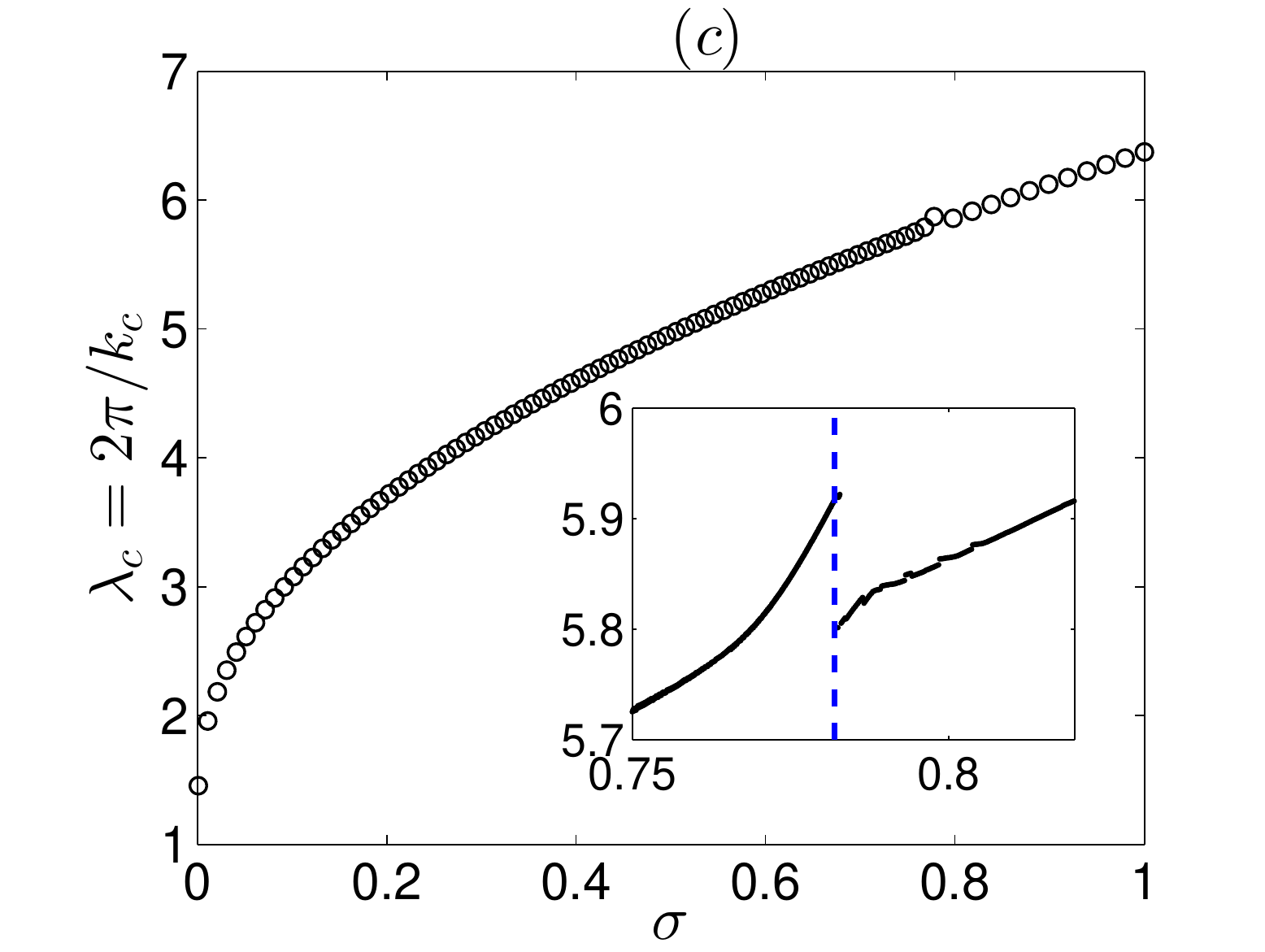}
\caption{(Color online) Stability diagram in the $(\theta,\varepsilon_{\text{dd}})$ plane as a function of $\sigma$ for $\theta=0$ (a) and $\theta=\pi/2$ (b) with the position of the phonon instability indicated by the dashed black line. The critical wavelength, $\lambda_{c}=2\pi/k_{c}$ at which the roton becomes unstable is shown as a function of $\sigma$ in (c). The inset shows a zoomed portion of this figure displaying the discontinuity in $\lambda_c$ at $\sigma_c\sim0.78$.}
\label{fig:evs}
\end{figure*}

Figure~\ref{fig:bdg} illustrates the dispersion relation corresponding to Eq.~\eqref{eqn:bdg}.  For certain parameters, the dispersion relation has the same structure as for the non-dipolar case (dot-dashed blue line): for low $k$ it is linear in $k$ (the phonon regime), changing to a quadratic form at higher $k$ (the free-particle regime).  The system is prone to a phonon (long wavelength $k_z \rightarrow 0$) instability. In non-dipolar homogeneous condensates, this arises when the mean-field van der Waals interactions become attractive, $g<0$.   Examining the small $k_z$ behavior of the dispersion $E_{k}=\hbar\omega_{k}$ relation given by Eq.~\eqref{eqn:bdg} yields
\begin{equation}
\omega_{k}=k_z\sqrt{\frac{1}{m}\bigg(-\frac{4n_0U_{0}}{3}+\frac{n_0g}{2\pi l_{\perp}^{2}}\bigg)}= k_z c_{s}
\label{eqn:pi}
\end{equation}
where $c_{s}$ is the speed of sound associated with this long-wavelength regime.
We identify the phonon instability as occurring when the bracketed term (which corresponds to the homogeneous chemical potential $\mu_0$) is less than zero.  This leads to imaginary frequencies, signifying the unstable exponential growth of long-wavelength perturbations.  

In other parameter regimes, the dipoles can change the form of the dispersion relation at intermediate $k$.  In certain instances, it causes a flattening off of the dispersion spectrum, as shown in Fig.~\ref{fig:bdg} (dashed black line), while in more extreme cases a minima can form in the dispersion relation at finite momentum, i.e. the roton.  An example is shown in Fig.~\ref{fig:bdg} (solid red line), with a pronounced dip appearing at $k_{z}\sim\xi^{-1}$.  The associated local maximum in the dispersion relation is termed the maxon.    If the roton minimum touches the zero-energy axis then the condensate undergoes the roton instability. The roton is predominantly driven by transverse (off-axis) effects of the dipole-dipole interaction, and becomes less pronounced in the one-dimensional limit ($\sigma \rightarrow 0$) \cite{lahaye_2009}.  

We identify the roton instability as follows. The expression \eqref{eqn:bdg} for $E_k^2(k)$  is differentiated with respect to $k_z$ and set equal to zero so as to identify the stationary points (which may correspond to the roton or the maxon). This is then combined with the dispersion relation \eqref{eqn:bdg} equated to zero (the maxon is automatically excluded from this result as it can never touch the zero energy axis).  With some manipulation, the critical wavenumber at which the roton touches zero energy is found to be
\begin{align}\nonumber
&k_{c}^{2}=-\frac{mn_{0}g}{\pi\hbar^2l_{\perp}^{2}\mu_0}\bigg\{\bigg[1+\frac{16\pi l_{\perp}^{3}U_{0}}{3g}\bigg]\\&-\sqrt{\bigg[1+\frac{16\pi l_{\perp}^{3}U_{0}}{3g}\bigg]^2-\frac{4\pi l_{\perp}^{2}\mu_0}{n_{0}g\sigma^{2}}\bigg[1-\frac{8\pi l_{\perp}^{3}U_{0}}{g}\bigg]}\bigg\}.
\label{eqn:kc}
\end{align}
While the above expression provides $k_c$ for known system parameters, we wish to predict the onset of the instability as a function of the dipole-dipole interaction strength $\varepsilon_{\text{dd}}$. We can eliminate $k_{c}$ by combining the above expression with a rearranged version of the dispersion relation \eqref{eqn:bdg} equated to zero: 
\begin{equation}
k_{c}^{2}+\frac{16n_{0}ml_{\perp}U_{0}}{\hbar^2}\mathcal{V}_{\text{1D}}\bigg(\frac{k_{c}^{2}l_{\perp}^{2}}{2}\bigg)+\frac{2gn_{0}m}{\pi l_{\perp}^{2}\hbar^2}=0 .
\label{eqn:td}
\end{equation}
These two equations can be solved iteratively to predict the critical $\varepsilon_{\text{dd}}$ for the roton instability to occur as a function of $\theta$ and $\sigma$, as will be presented below.



In Figure \ref{fig:evt} we map out stability diagrams in the $(\theta,\varepsilon_{\text{dd}})$ plane, showing the regions of roton instability (shaded red) and phonon instability (shaded blue).  To give insight into the role of the transverse condensate width, three values are considered: (a) $\sigma=0.1$, (b) $0.5$ and (c) $1$.  For each value, we  distinguish between the $g>0$ (left column) and $g<0$ (right column) cases. The phonon instability is independent of $\sigma$ throughout, and intuitively follows from the play-off between the van der Waals interactions and the dipolar contribution to the contact interactions. Consider, for example, the case of repulsive vdW interactions.  For $\theta=0$ the dipoles lie in the attractive end-to-end configuration, and  when the dipoles are sufficiently strong ($\varepsilon_{\rm dd}>1$) they can overwhelm the repulsive vdW interactions, inducing phonon instability.  Conversely, for $\theta=\pi/2$ the dipoles are side-by-side; conventionally this is a repulsive configuration, but in the regime of anti-dipoles ($C_{\rm dd}<0$) this configuration is attractive, and induces phonon instability when the anti-dipoles are sufficiently strong ($\varepsilon_{\rm dd} < -2$).   

The regions of roton instability are sensitive to $\sigma$.  For low $\sigma$ (cases (a) and (b) in Figure \ref{fig:evt}), the roton instability arises only for attractive vdWs interactions ($g<0$).  Deep in the 1D regime (case (a)) the roton instability arises only in a narrow band in the $(\theta,\varepsilon_{\text{dd}})$ plane; as $\sigma$ is increased this band expands (case (b)).  However once a critical value ($\sigma_{\text{crit}}\sim0.8$) is exceeded, as per row (c), the roton instability shifts instead to appearing only for repulsive vdW interactions ($g>0$). 
The value of $\sigma_{\text{crit}}$ does not depend on the angle $\theta$. 

To further explore the shift of the roton instability from $g<0$ to $g>0$ for increasing $\sigma$,  Fig.~\ref{fig:evs} depicts the roton instability in the $(\sigma,\varepsilon_{\text{dd}})$ plane for the extreme polarization angles,  (a) $\theta=0$ and (b) $\theta=\pi/2$.  In both cases it is seen that in the quasi-1D limit ($\sigma\ll1$) the roton instability band narrows, approaching the onset of the phonon instability (horizontal dashed line). However as $\sigma$ is gradually increased, the roton instability undergoes a change of sign at the critical value $\sigma=\sigma_{\text{crit}}$. Figure~\ref{fig:evs} (c) shows the critical wavelength defined as $\lambda_{c}=2\pi/k_{c}$ plotted as a function of $\sigma$. It is seen that $\lambda_{c}$ is monotonically increasing, and is always greater than $\sigma$, indicating that our one-dimensional analysis is valid. The inset to Fig.~\ref{fig:evs} (c) shows a zoomed in portion of this graph centered around $\sigma=\sigma_c$, and clearly shows the discontinuity in $\lambda_c$, indicated by the dashed blue line.

\begin{figure}[t]
\centering
\includegraphics[width=1\columnwidth]{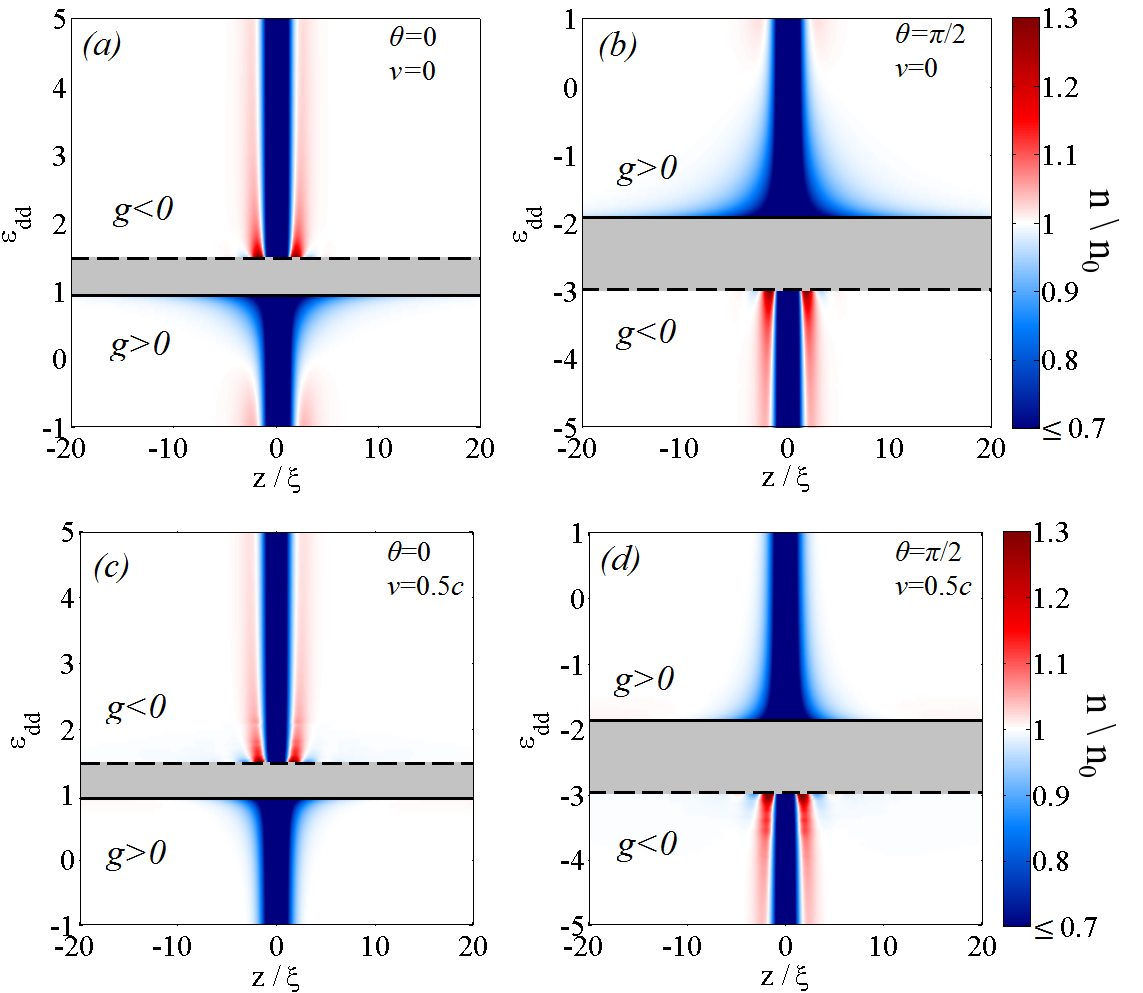}
\caption{(Color online) Density profiles $n(z)$ for (a) $v=0$ and $\theta=0$, (b) $v=0$ and $\theta=\pi/2$, (c) $v=0.5c$ and $\theta=0$ and (d) $v=0.5c$ and $\theta=\pi/2$, as a function of $\varepsilon_{\rm dd}$ with $\sigma=0.1$. The band (grey) of instability is bounded by the onset of the roton (dashed) and phonon (solid) instabilities. Note that either side of this unstable band, the system is stable for only $g>0$ or $g<0$, as indicated.  }\label{fig:1sol}
\end{figure}

\section{Dark Soliton Solutions\label{sec:dds}}
\subsection{Non-Dipolar Dark Soliton Solutions}
In the absence of dipolar interactions ($\varepsilon_{\rm dd}=0$) and for repulsive $s$-wave interactions ($g>0$) it is well-known that the 1D GPE is integrable and supports dark soliton solutions of the form \cite{zakarov_1973,kivshar_1998},
\begin{equation}
\psi_{\rm s}(z,t)=\sqrt{n_0}\left[ \beta \tanh \frac{\beta (z-vt)}{\xi}+i \frac{v}{c} \right]e^{-i \mu t/\hbar}.
\label{eqn:ds}
\end{equation}
Here $\beta=\sqrt{1-v^2/c^2}$, where $v$ is the velocity of the soliton, $c$ is the sound speed in the condensate, and the dark soliton's centre of mass is initially placed at the origin. The family of solitons commonly feature a density depression and a phase slip, with the depression density, $n_d$ and total phase slip $S$ related to the soliton speed, $v$, via $v/c=\sqrt{1-(n_d/n_0)}=\cos(S/2)$. The $v=0$ ``black'' soliton ($z=0$) has zero density at its centre and a $\pi$ phase slip, while the $v=c$ soliton is indistinguishable from the background density. Since dark solitons deplete the density profile, they are analogous to particles with negative mass \cite{busch_2000}.
\subsection{Dipolar Dark Soliton Solutions}
\begin{figure}[b]
\centering
\includegraphics[width=0.9\columnwidth]{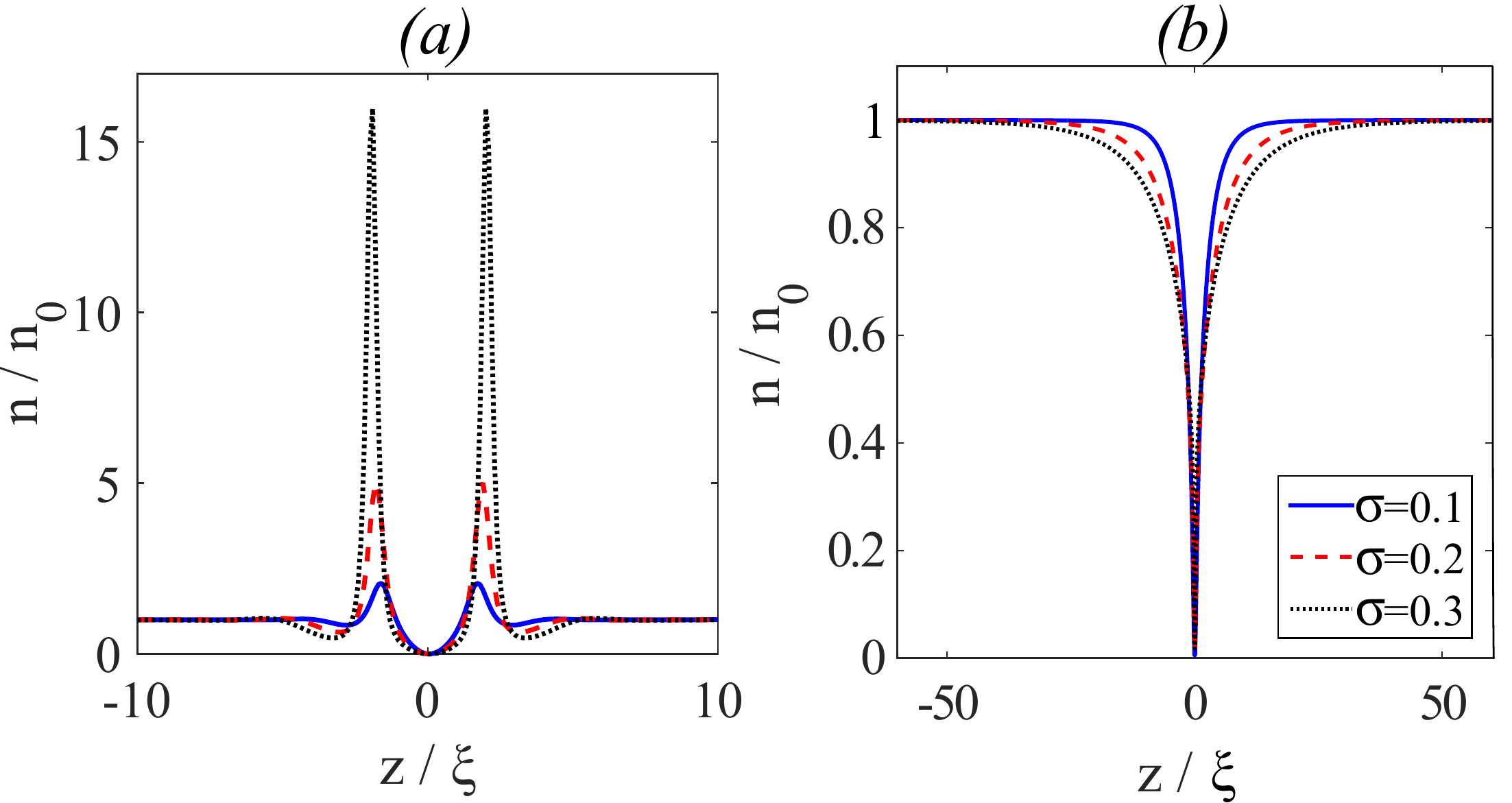}
\caption{(Color online) Density profiles of the black ($v=0$) soliton close to the (a) roton instability and the phonon instability (b), for 3 values of $\sigma$.  Since the roton instability shifts with $\sigma$, we consider different values of $\varepsilon_{\text{dd}}$ in (a): $\varepsilon_{\text{dd}}=1.46$ for $\sigma=0.1$, $\varepsilon_{\text{dd}}=2.68$ for $\sigma=0.2$, and $\varepsilon_{\text{dd}}=7.60$ for $\sigma=0.3$.  In (b) we take $\varepsilon_{\text{dd}}=0.95$ throughout.}\label{fig:sigma_dep}
\end{figure}

We now set about exploring the dipolar dark soliton solutions across the important parameters of soliton speed $v$, polarization angle $\theta$, dipolar strength $\varepsilon_{\text{dd}}$, and condensate width $\sigma$. We consider the dark solitons as stationary states in the moving frame, and obtain these solutions by numerically solving the 1D dipolar GPE in the moving frame.  The details of this approach can be found in Appendix A.
Figure~\ref{fig:1sol}(a,b) plots the spatial density profile $n(z)$ of the black ($v=0$) soliton solutions as a function of $\varepsilon_{\text{dd}}$, for the extreme polarization angles of (a) $\theta=0$ and (b) $\theta=\pi/2$. The former represents the typical behavior for all solutions in the range $\theta<\theta_{\rm m}$, and the latter shows the converse. Figure~\ref{fig:1sol}(c,d) shows the corresponding plot for a finite speed case, $v=0.5c$.  The grey bands represent the range of $\varepsilon_{\text{dd}}$ for which no stable condensate exists, consistent with the corresponding stability diagrams presented in row $(a)$ of Fig. \ref{fig:evt}.  Away from the unstable region, the density profiles resemble the $\tanh^2$-density of the conventional dark solitons, with a width of the order of the healing length $\xi$ (which should be noted is itself a function of $\varepsilon_{\rm dd}$ and $\theta$).  Near to the phonon instability (solid lines) the density profile diverges in width; this is due to a cancellation between the local interactions arising from the explicit van der Waals interactions and the implicit local contribution to the DDIs, with a similar effect seen for vortices in 2D dipolar condensates \cite{mulkerin_2013}. Meanwhile, as the roton instability (dashed lines) is approached, density ripples form symmetrically around the soliton, decaying as they recede from the core.  For the cases shown in Fig.~\ref{fig:1sol}, the ripples can rise to twice the background density with the most prominent parts being the two dominant lobes either side of the dark soliton (see also Fig. \ref{fig:sigma_dep}).  They arise due to the soliton state mixing with the roton, an effect akin to that predicted for vortices in 2D~ \cite{yi_2001,wilson_2008,mulkerin_2013}. The ripples can be understood from an energetic point of view by noticing that they occur when the dipolar interactions are repulsive along the axial direction meaning that the system can lower its energy by placing more dipoles near to the empty core. The repulsive dipolar interaction due to the lobes in turn causes a density reduction next to them, then another peak, and so on. The ripples are thus a direct effect of the long range nature of the dipolar interactions.  Despite the density modulations, the soliton depth still follows the relation $n_d/n_0=1-v^2/c^2$, familiar from non-dipolar dark solitons. We also find that the density modulations are slightly enhanced for slower solitons.

As previously discussed in Sec.~\ref{sec:inst}, the roton and its instability is sensitive to the condensate width, $\sigma$. To determine the effect of $\sigma$ on dark solitons, Fig. \ref{fig:sigma_dep} compares the black soliton solution close to the (a) roton instability and (b) phonon instability, for different values of $\sigma$.  Note that we maintain $\sigma< 1$ throughout so to satisfy the governing criteria for a quasi-1D condensate (see Section \ref{sec:mf1d}).  As $\sigma$ increases the system becomes less ``1D" in nature and the effects of dipolar interactions become more pronounced.  At the roton instability, the density ripples grow rapidly with $\sigma$, becoming as large as $15n_0$ for $\sigma=0.3$ (dotted black line).  The length scale of the ripples also increases with $\sigma$; this is consistent with the earlier homogeneous analysis, which showed the roton wavelength to increase with $\sigma$ (Fig.~\ref{fig:evs} $(c)$). Meanwhile, at the phonon instability the soliton has a funnel shaped density profile, which widens with $\sigma$.  Having explored the dependency of $\sigma$ we will employ $\sigma=0.1$ for the remainder of our work (the relevant homogeneous stability diagrams being shown in row $(a)$ of Fig. \ref{fig:evt}).

We briefly comment on the manifestation of the phonon and roton instabilities on the soliton solutions. Imagine crossing the phonon instability threshold, from the stable side to the unstable side. The net interactions switch from repulsive to attractive, such that dark solitons are no longer stable.  At the same time the background condensate undergoes a modulational instability, as per the non-dipolar attractive condensate \cite{brand}, and fragments into bright soliton-like structures (the stable structures under net attractive interactions). Next consider crossing the roton instability, again from stable to unstable. The ripples surrounding the dark soliton grow and we find the BEC eventually collapses. However, in both cases, the sharp growth in density means that higher-order effects such as three-body losses \cite{lahaye_2008} and quantum fluctuations \cite{barbut_2016} become important. Such physics is not contained within our dipolar GPE and is beyond the scope of this work.


\subsection{Integrals of Motion}
The family of non-dipolar dark solitons [Eq.~(\ref{eqn:ds})] possess an infinite number of integrals of motion (viz. conserved quantities).  The first three of these have a clear physical meaning: the soliton normalization, momentum, and energy \cite{kivshar_1998,frantzeskakis_2010}. In order to calculate finite values for each of these quantities, we compute their {\it renormalized} versions, that is, the difference between the quantity in the presence and absence of the soliton. In this subsection we generalize these quantities for dipolar dark solitons and explore their dependence on the dipolar parameters. 

The renormalized norm and momentum of the dark soliton are defined as
\begin{eqnarray}
N_{\text{sol}}&=&\int_{-\infty}^\infty\di z~(n_0-|\psi|^2),
\label{eqn:sol_norm}
\\
P_{\text{sol}}&=&\frac{i\hbar}{2}\int_{-\infty}^\infty\di z\bigg[\psi\frac{\partial\psi^{*}}{\partial z}-\psi^{*}\frac{\partial\psi}{\partial z}\bigg]\bigg(1-\frac{n_0}{|\psi|^2}\bigg),
\label{eqn:sol_mom}
\end{eqnarray}
where $n_0$ is the homogeneous background density.  The renormalized energy must explicitly include the contribution from the dipoles.  To calculate this we follow a similar approach to Ref. \cite{pethick_2002}.  According to Eq. (\ref{eqn:en}) the energy of a homogeneous system of length $L$ is $E_0=(gn_0^2/4\pi l_\perp^2+\Phi_{0}n_0/2)L$, where $\Phi_0$ is the homogeneous dipolar potential.  In order to make a direct comparison between a homogeneous system and a system containing a soliton the quantity $E-\mu N$ must be considered, so as to account for the different particle number $N$ between the two systems.  Thus the renormalized soliton energy 
can be written as
\begin{eqnarray}
E_\text{sol}=\int^{\infty}_{-\infty} \di z \Bigg[\frac{\hbar^2}{2m}\bigg|\frac{\partial\psi}{\partial z}\bigg|^2+\frac{g}{4\pi l_\perp^2}(|\psi|^2-n_0)^2 \nonumber \\
+\left(\frac12\Phi_{\text{dd}}-\Phi_0\right)|\psi|^2+\frac12\Phi_{0}n_0\Bigg].
\label{eqn:soli_energy}
\end{eqnarray}
In the absence of a soliton, for which $\Phi_{\text{dd}}=\Phi_0$ and $|\psi|^2=n_0$, this expression correctly reduces to zero.

In the absence of dipoles and using the soliton solution (\ref{eqn:ds}), the renormalized norm, momentum and energy follow as
\begin{subequations}
\begin{eqnarray}
N_\text{sol}^0&=&2\xi n_0\beta,
\label{eqn:ndsn}
\\ P_\text{sol}^0&=&-\frac{2\hbar n_0v\beta}{c}+2\hbar n_0\arctan\left(\frac{c\beta}{v}\right),
\label{eqn:ndsp}
\\E_\text{sol}^0&=&\frac43n_0\hbar c\beta^{3},
\label{eqn:ndse}
\end{eqnarray}
\end{subequations}
where, recall, $\beta = \sqrt{1-v^2/c^2}$. Meanwhile the effective mass of the non-dipolar dark soliton is found from the relation $m_{\star}^{0}=\partial P_{\text{sol}}^{0}/\partial v$, and is given by
\begin{equation}
m_{\star}^{0}=-\frac{4\hbar n_0\beta}{c}.
\end{equation}

\begin{figure}[t]
\centering
\includegraphics[scale=0.28]{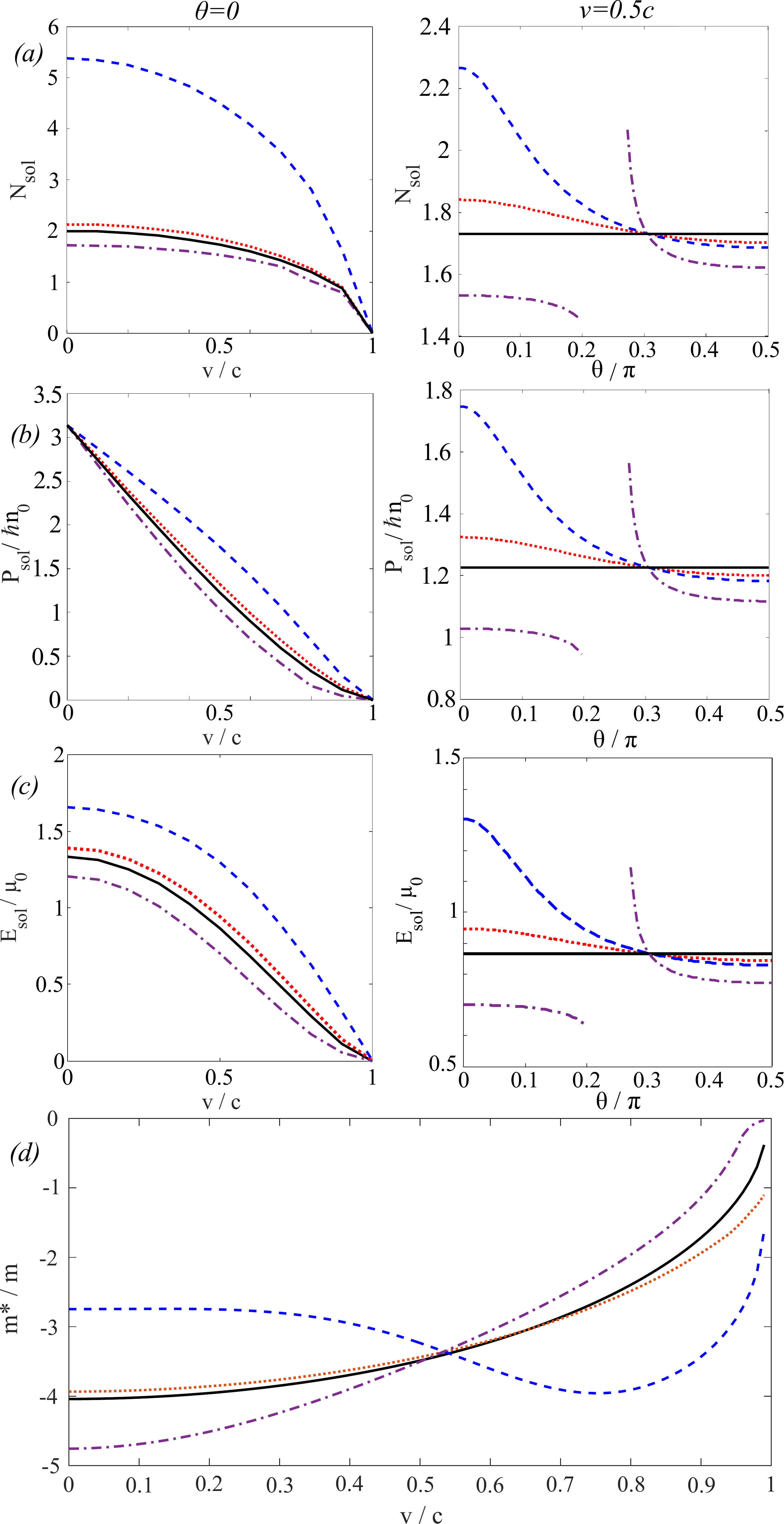}
\caption{(Color online) Three integrals of motion, the soliton $(a)$ norm, $(b)$ momentum, $(c)$ energy, as a function of $v/c$ (left column, with $\theta=0$) and $\theta$ (right column, with $v=0.5c$). $(d)$ The soliton effective mass $m^{\star}$ as a function of soliton velocity. All plots contain four lines showing the non-dipolar solution (solid black), $\varepsilon_{\text{dd}}=0.4$ (dotted red), $\varepsilon_{\text{dd}}=0.8$ (dashed blue), and $\varepsilon_{\text{dd}}=5$ (purple dot-dashed).}\label{fig:iom}
\end{figure}

\begin{figure*}
\centering
\includegraphics[scale=0.35]{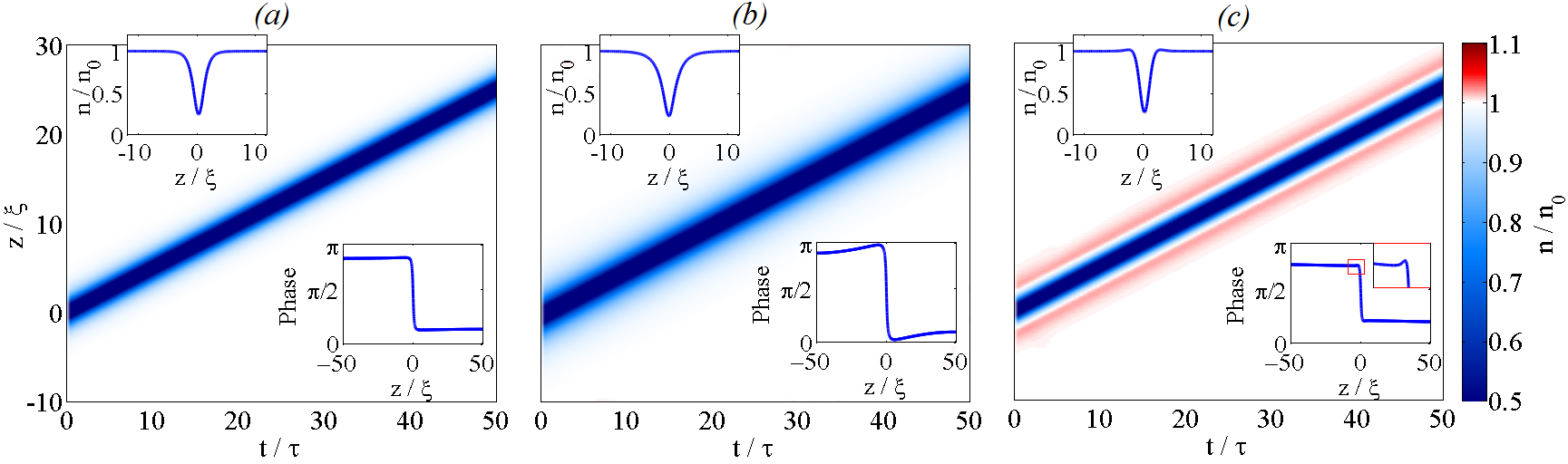}
\caption{(Color online) Single dipolar dark solitons propagating with unchanging form with speed $v=0.5c$ (and $\theta=0$) for (a) $\varepsilon_{\text{dd}}=0.4$ ($^{168}$Er parameters),  (b) $\varepsilon_{\text{dd}}=0.8$ (close to the phonon instability), and (c) $\varepsilon_{\text{dd}}=5$ (close to the roton instability). Top left insets show the soliton density profile and bottom right insets show the soliton phase profile.}\label{fig:soli_move}
\end{figure*}

We evaluate the norm, momentum and energy of the dipolar solitons numerically according to Eqs. (\ref{eqn:sol_norm}-\ref{eqn:soli_energy}). In the left rows of Fig. \ref{fig:iom}~(a-c) these conserved quantities are plotted as a function of the soliton speed for three values of $\varepsilon_{\rm dd}$, with the polarization angle fixed to $\theta=0$.  Note that these three values of $\varepsilon_{\rm dd}$ correspond to $^{168}$Er parameters ($\varepsilon_{\rm dd}=0.4$) and near to the phonon  ($\varepsilon_{\rm dd}=0.8$) and roton instabilities  ($\varepsilon_{\rm dd}=5$). Throughout, the dipolar results show the same qualitative structure as the non-dipolar result (solid black line), but can be quite different quantitatively. The soliton norm $N_{\rm sol}$ [Fig. \ref{fig:iom} $(a)$, left] decreases monotonically with speed due to the reduction in the soliton depth for larger speeds, becoming zero for $v=c$ when the soliton is indistinguishable from the background. The soliton momentum $P_{\rm sol}$ [Fig. \ref{fig:iom} $(b)$, left] also decreases monotonically with $v$, due to the decreasing norm and decreasing phase gradient (the phase gradient determines the local fluid velocity according to $u(z)=(\hbar/m)\partial_z S(z)$, where $S(z)$ is the phase profile)  across the soliton.  It has a universal value for $v=0$ of $P_{\text{sol}}=\pi \hbar n_0$ due to the $\pi$-phase-step profile of the $v=0$ solitons.  The momentum becomes zero for $v=c$ when the norm and phase gradient reach zero. The dipoles have the greatest effect on $P_{\rm sol}$ for intermediate velocities.  Finally, the soliton energy $E_{\rm sol}$ [Fig. \ref{fig:iom} $(c)$, left]  also decreases with $v$, associated with the decreasing interaction energy as the soliton gets increasingly shallow and the decreasing kinetic energy due to the reduced density and phase gradients, and at $v=c$ $E_{\rm sol}=0$.

Close to the phonon instability (blue dashed lines) the integrals of motion tend to be larger than the non-dipolar case.  This can be related to the wide funnel-shaped profile which develops close to this instability.  This vastly increases the effective volume of the soliton core, i.e. the norm.  This in turn raises the momentum and energy (in the latter case, due primarily to the increased interaction energy associated with the larger density depletion). The momentum and energy are also modified by density gradients. Meanwhile, close to the roton instability (purple dot-dashed line) the integrals of motion tend to be smaller.  The density ripples which form here act to reduce the effective volume of the soliton core, which reduces the momentum and energy relative to the non-dipolar case.  

Finally, in the right rows of Fig. \ref{fig:iom}~(a-c) the conserved quantities are plotted as a function of the polarization angle for the three values of $\varepsilon_{\rm dd}$, while keeping the soliton speed fixed ($v=0.5c$).  The non-dipolar result is constant in each plot.  What is particularly prominent is that the $\theta$ dependence is the same for all 3 integrals of motion, with only the scale changing.  At the magic angle, $\theta_{\rm m}\approx 0.3\pi$, the integrals equal the non-dipolar result, due to the vanishing of the dipolar potential here.  Note that the gap in the curve for $\varepsilon_{\rm dd}=5$ is due to the absence of stable solutions here, consistent with the stability diagram in Fig. \ref{fig:evt} $(a)$. 

The effective mass of the soliton, defined as $m^\star=\partial P_{\text{sol}}/\partial v$, is shown in  Fig. \ref{fig:iom}~(d). The effective mass is negative throughout, tending towards zero effective mass when $v=c$, as expected.  For most cases the effective mass increases monotonically with $v$.  However, close to phonon instability it has a unusual form, being approximately constant for $v/c\lesssim 0.4$ and decreasing to a local minimum at $v/c\approx 0.75$ $m^{\star}/m$.  

\section{Dynamics of Dipolar Dark Solitons\label{sec:dsint}}

In this section we explore the rudimentary dynamics of the dipolar dark solitons.  In particular we seek to establish their soliton-like nature.  We will approach this by reference to the general definition of a soliton given by Johnson and Drazin \cite{drazin_1989}, which is of three key properties: (i) permanent form, (ii) localized within a region of space, and (iii) emergence from collisions  unchanged, barring a phase shift.  The results presented here are based on numerical propagation of the 1D (lab-frame)  dipolar GPE using the Crank-Nicolson method, using a suitable initial condition featuring soliton solutions obtained from the BCGM method. 
\subsection{Propagation}

Figure \ref{fig:soli_move} shows the evolution of a $v=0.5c$ dipolar dark soliton (with fixed $\theta=0$) for three values of $\varepsilon_{\rm dd}$: $0.4$ (corresponding to $^{168}$Er), $0.8$ (close to the phonon instability) and $5$ (close to the roton instability).  Insets show the density and phase profiles across the soliton.  For all three cases, the soliton maintains a permanent and localized form, with no radiative losses.  It also undergoes centre-of-mass translation at the expected speed.   As such, these states satisfy the soliton criteria (i) and (ii) above.  It is also worth observing the phase profile across the soliton.  For $\varepsilon_{\rm dd}=0.4$ [Fig. \ref{fig:soli_move}(a)], the phase profile is practically identical to that of the non-dipolar dark soliton, with a tanh-shaped step which relaxes to the asymptotic value over a short length scale of the order of the healing length.  Close to the instabilities, the phase relaxes over a much larger length scale, of  around $\sim50\xi$ close to the phonon  instability [Fig. \ref{fig:soli_move}(b)] and around $400\xi$ close to the roton instability [Fig. \ref{fig:soli_move}(c)].  In all cases the total asymptotic phase slip is the same as for the non-dipolar soliton (this is not directly evident from the inset in (c) due to the limited length range of this plot).  Close to the instabilities the phase profile also features distinctive prominences.  At the phonon instability these are broad, while at the roton instability they are of order of the roton length scale.  



\subsection{Collisions}
\begin{figure}[t]
\centering
\includegraphics[scale=0.28]{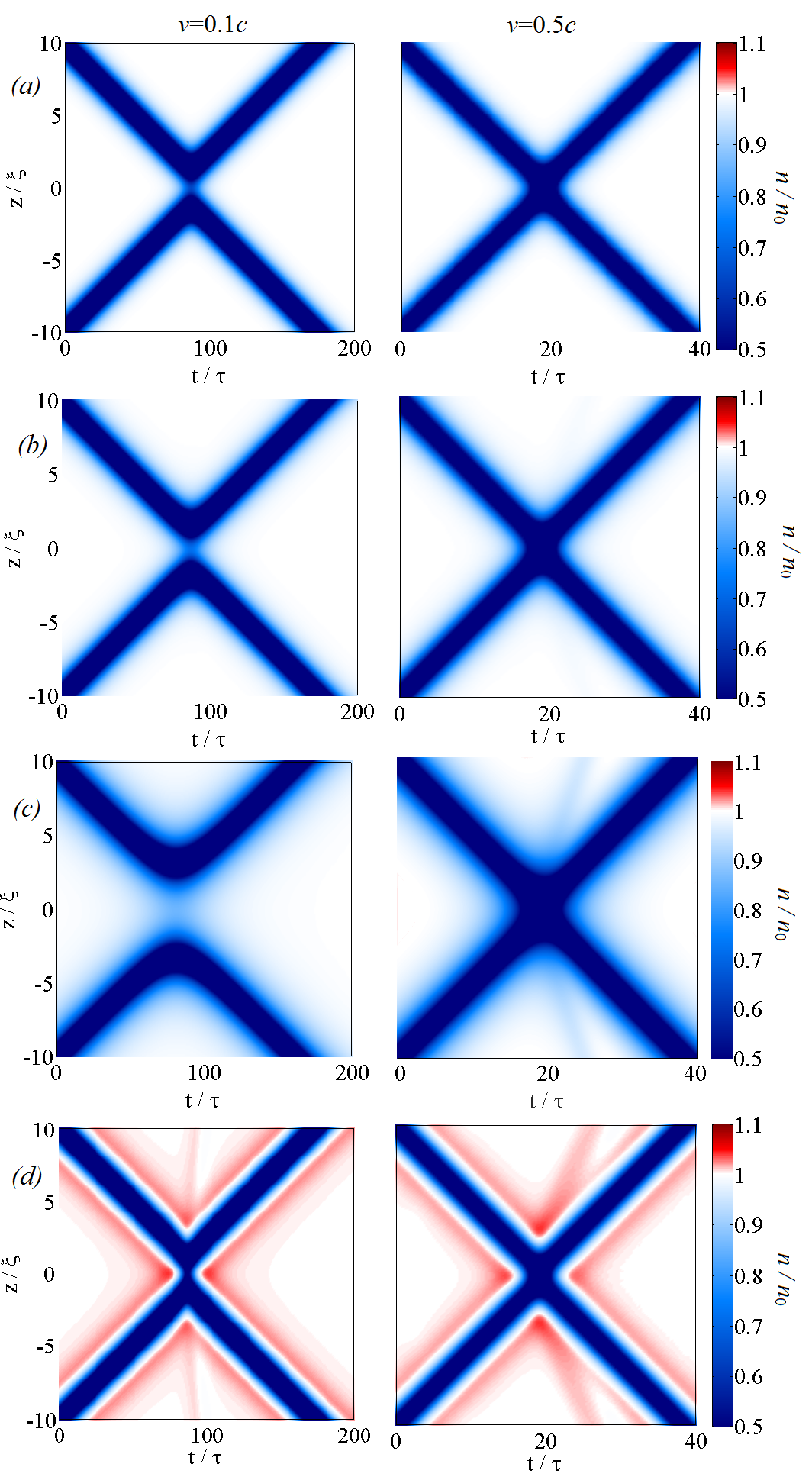}
\caption{(Color online) Collisions of two dark solitons at low speed ($v=\pm 0.1c$, left panels) and higher speed ($v=\pm 0.5c$, right panels) for $(a)$ $\varepsilon_{\text{dd}}=0$, $(b)$ $\varepsilon_{\text{dd}}=0.4$, $(c)$ $\varepsilon_{\text{dd}}=0.8$ and $(d)$ $\varepsilon_{\text{dd}}=5$. The polarization angle is taken to be $\theta=0$.}\label{fig:collision}
\end{figure}
In non-dipolar systems the interaction between multiple dark solitons has been experimentally observed and theoretically studied \cite{weller_2008,theocharis_2010}. In a symmetric collision for solitons satisfying $0< v/c<\frac{1}{2}$ the $s$-wave interactions create a repulsive force causing the solitons to appear to reflect at short distance. For velocities satisfying $\frac{1}{2}\leq v/c<1$ the solitons appear to pass through each other.  In both cases the outgoing solitons are unchanged from the incoming form, barring a phase shift. In the presence of dipolar interactions this behavior is modified \cite{pawlowski_2015,bland_2015}. In Ref. \cite{bland_2015} the effect of polarization angle on the collisions was explored, revealing an additional non-local repulsion or attraction on the soliton-soliton interaction due to the dipolar interactions.  The latter case, in combination with the conventional short-range repulsion, was shown to support bound states.  Here we explore the soliton collisions further, exploring the role of the experimentally-tunable interaction parameter $\varepsilon_{\rm dd}$ and the soliton speed.  We investigate the effect of the dipolar contribution near to the instabilities, for a system comprised of $^{168}$Er atoms \cite{aikawa_2012}, with polarization $\theta=0$.

Figure \ref{fig:collision} presents a series of dark soliton collisions for (left column) slow incoming solitons ($v=\pm 0.1c$) and (right column) faster solitons ($v = \pm 0.5c$), and for various values of $\varepsilon_{\rm dd}$.  For comparison, the non-dipolar collisions are also shown (row (a)); these confirm the apparent bouncing for low collisional speeds and apparent transmission for higher speeds.  For relatively weak dipolar interactions (row (b), $\varepsilon_{\rm dd}=0.4$) the soliton collisions are virtually indistinguishable from the non-dipolar case, with a short-range-dominated interaction and the solitons emerging unscathed.  

For stronger dipolar interactions (row (c), $\varepsilon_{\rm dd}=0.8$), the dipolar interactions have a noticeable effect on the soliton interactions, particularly at low speeds.  Here the solitons appear to bounce at a considerably greater separation than the non-dipolar case.  This effect can be directly related to the dipolar interactions. Since $\theta=0$ and $C_{\rm dd}>0$ for this case, the dipoles lie end-to-end and attract each other.  However, the dark soliton is a region of depleted dipoles, and can be interpreted as a positive density of anti-dipoles \cite{klawunn_2008,bland_2015} which repel each other.  This repulsive contribution to the dark soliton interaction will arise whenever the dipoles are net repulsive, i.e. when $C_{\rm dd}>0$ with $\theta<\theta_m$ or when $C_{\rm dd}<0$ with $\theta>\theta_m$.  The repulsive nature of the collision becomes washed out at higher incoming speeds (row (c), right panel).  While the soliton collision is stable at low speed (left panel), at high speed the collision is inelastic, with energy lost from the solitons via the emission of sound waves (visible as bands propagating away from the collision at the speed of sound).  

The case  of $\varepsilon_{\rm dd}=5$ (row (d)) instead has $C_{\rm dd}<0$, i.e. repulsive dipoles.  This in turn leads to an attractive contribution to the soliton interaction, and this is clearly observed in the corresponding soliton collisions.  Note the distinctive sharp ``pinching'' of the solitons during their collision at low speed.  More generally, this attractive contribution arises whenever the dipolar interactions are net attractive, i.e.  when $C_{\rm dd}<0$ with $\theta<\theta_m$ or when $C_{\rm dd}>0$ with $\theta>\theta_m$.  Here, for both low and higher incoming speeds, the collisions are inelastic through sound emission.  

Note that when the dipoles are polarized at the magic angle $\theta_m$ the dynamics are equivalent to the non-dipolar case \cite{bland_2015}.


Away from the phonon and roton instabilities, the solitons collide elastically, and emerge unscathed from the collision. This satisfies the third soliton criteria outlined above.  However, close to the instabilities, the collisions become dissipative, with sound being radiated away.  This is particularly prevalent for higher speed collisions.  We note, however, that the energy dissipated into sound waves during a single collision is typically very small, for example, in the maximally-dissipative case presented in Fig. \ref{fig:collision} ((d), right panel), the energy loss is $\sim10^{-3}\% E_{\rm sol}$.  






\section{Conclusions\label{sec:conc}}
In this work the family of dark solitons supported in a quasi-one-dimensional dipolar Bose-Einstein condensate were studied.  A bi-conjugate gradient method was implemented to numerically obtain these non-trivial solitons as stationary solutions in the moving frame, as a function of the dipole-dipole interaction strength ($\varepsilon_{\text{dd}}$), the polarization angle $\theta$ and the soliton speed. The phonon and roton instabilities of the system play a key role in modifying the density and phase profiles of the solitons, which can deviate significantly from the non-dipolar form in these regimes. The dipolar dark solitons were characterized in terms of their integrals of motion (norm, momentum and energy).  Due to the modified profiles in the presence of dipolar interactions, these quantities differ from their non-dipolar form, particularly so close to the instabilities.   The prominent role of the phonon and roton instabilities in the soliton solutions motivated a detailed and general analytical treatment of these effects in the quasi-1D dipolar condensate.  This, in particular, revealed the sensitivity of the roton to the transverse condensate size, $\sigma$, and the increase in the roton length scale as the dimensionality crossover, $\sigma \sim 1$, is approached. 

In isolation, the solitons propagate with unchanged form throughout the parameter space.  Away from the instabilities their collisions are elastic, but become dissipative via emission of sound waves close to the instabilities.  Thus, close to the instabilities these structures deviate from solitons in a strict sense, although it should be noted that the energy dissipated in a single collision is very small.  


Data supporting this publication is openly available under an Open Data Commons Open Database License \cite{data}.

\section{Acknowledgements}
This work was supported by the Engineering and Physical Sciences Research Council (Grant No. EP/M005127/1).  D. O. also acknowledges support from the Natural Sciences and Engineering Research Council (NSERC) of Canada.  We thank Nick Proukakis and Joachim Brand for discussions.

\appendix
\section{Numerical Approach to the Dipolar Dark Soliton Solutions}\label{app:a}

In this appendix, we describe how the soliton solutions in the main body of the paper were obtained.  We consider the dark solitons as stationary solutions of the GPE in the moving frame.  We obtain them by numerically solving the moving-frame time-independent GPE, $(\hat{\mathcal{H}} + v\hat{p}_{z}-\mu)\phi=0$, where $\hat{p}_{z}=-i\hbar\partial/{\partial z}$ defines the momentum operator in the $z$-direction.  This is performed using the bi-conjugate gradient method \cite{Press_Flannery_90} (BCGM).  This technique has been used to obtain moving-frame vortex solutions in the 2D/3D GPE \cite{Winiecki_McCann_99,winiecki}. 
 It is worth noting that this approach provides the true dipolar dark soliton solutions for arbitrary speed. In contrast, the approach to finding soliton solutions based on imaginary time propagation of the GPE \cite{pawlowski_2015} requires {\it a priori} knowledge of the soliton phase, and so is only capable of obtaining black ($v=0$) soliton solutions, for which the phase profile is known to be a step function of amplitude $\pi$.

The BCGM is an iterative method based on the Newton-Raphson method for finding roots of equations. For a function $f(x)$, Newton's method uses an initial guess, $x^{(1)}$, and the following iteration 
\begin{equation}
x^{(p+1)}=x^{(p)}-\frac{f(x^{(p)})}{f'(x^{(p)})}\label{eq_newton}
\end{equation}
to minimise the Taylor expansion $f(x^{(p+1)})\approx f(x^{(p)})+f'(x^{(p)})(x^{(p+1)}-x^{(p)})=0$ at step $p$. Generalizing this to a system of $N$ coupled equations, $\mathbf{f(x)=0}$ (here $\mathbf{x}$ denotes the vector $\{x_u\}_{u=1,...,N}$ and $\mathbf{f(x)}$ the vector of functions $\{f_u\}_{u=1,...,N}$), the same minimization procedure can be applied.  

Defining $\mathbf{f}(\phi)=(\hat{\mathcal{H}} + v\hat{p}_{z} - \mu)\phi$, the linearized system of equations we seek to solve is
\begin{equation}
f_u(\phi^{(p+1)})\approx f_u(\phi^{(p)})+\sum_{v=1}^{N}J_{u,v}\delta\phi_v\approx0,
\label{eqn:lin}
\end{equation}
where $J_{u,v}=\partial f_u(\phi^{(p)})/\partial\phi_v^{(p)}$ defines the elements of the Jacobian matrix and $\delta \phi=\phi^{(p+1)}-\phi^{(p)}$.

Position $z$ is discretized onto a grid $z_i$, with $i=1,..., \mathcal{N}$ and grid spacing $\Delta z$. Similarly, the spatial wave function is denoted $\phi_j$ ($j=1,..., \mathcal{N}$).  We further define the real and imaginary parts of $\phi_j$ as $\phi_{j,0}=\text{Re}[\phi(z_j)]$ and $\phi_{j,1}=\text{Im}[\phi(z_j)]$.  The discretized version of the function $\bf{f}$ can then be written down in terms of its composite real and imaginary parts as
\begin{eqnarray}
f_{j,r} =&-&\frac{\hbar^2}{2m}\frac{\phi_{j-1,r}-2\phi_{j,r}+\phi_{j+1,r}}{(\Delta z)^2}\nonumber \\ 
&+&\Bigg[\frac{g}{2\pi l_\perp^2}(\phi_{j,0}^2+\phi_{j,1}^2)-\Phi_{\text{dd}}^{j,r}-\mu\Bigg]\phi_{j,r} \nonumber \\
&+&(2r-1)v\hbar\frac{\phi_{j+1,1-r}-\phi_{j-1,1-r}}{2\Delta z}=0,
\label{eq:bcgm}
\end{eqnarray}
where the spatial derivatives have been evaluated through the finite-difference scheme.  The computation of the dipolar potential $\Phi_{\text{dd}}^{j,r}$ is handled via the convolution theorem as
\begin{eqnarray}
\Phi_{\text{dd}}^{j,r}=\mathcal{F}^{-1}[\mathcal{F}[U_\text{1D}(z_j)]\mathcal{F}[|\phi_{j,r}|^2]].
\end{eqnarray}

The Jacobian $J_{u,v}$ in Eq. (\ref{eqn:lin}) is formed as the discrete functional derivative of $f_{j,r}$ with respect to $\phi_{k,s}$. Making use of the relation $\partial \phi_{j,r}/\partial \phi_{k,s}=\delta_{j,k}\delta_{r,s}$, we obtain
\begin{eqnarray}
\frac{\partial f_{j,r}}{\partial \phi_{k,s}} =&-&\frac{\hbar^2}{2m}\frac{\delta_{j+1,k}-2\delta_{j,k}+\delta_{j-1,k}}{(\Delta z)^2}\delta_{r,s}\nonumber \\ 
&+&\delta_{j,k}\delta_{r,s}\Bigg[\frac{g}{2\pi l_\perp^2}(\phi_{j,0}^2+\phi_{j,1}^2)-\Phi_{\text{dd}}^{j,r}-\mu\Bigg] \nonumber \\
&+&2\phi_{j,r}\phi_{k,s}\left[\frac{g}{2\pi l_\perp^2}\delta_{j,k}-U_{\text{1D}}(|z_j-z_k|)\Delta z\right] \nonumber \\
&+&(2r-1)v\hbar\delta_{1-r,s}\frac{\delta_{j+1,k}-\delta_{j-1,k}}{2\Delta z}.
\label{eq:jac}
\end{eqnarray}

Numerical implementation was handled in MATLAB with the function \texttt{bicgstab}. In practice, it is convenient to concatenate the real and imaginary components of \textbf{f} into a single vector of length $2\mathcal{N}$, and similarly for $\phi$. \textbf{J} is then of size $2\mathcal{N}\times 2\mathcal{N}$. Taking the non-dipolar soliton solution, as defined by Eq. (\ref{eqn:ds}), as the initial guess for $\phi$ (centered at the origin), we find that the BCGM robustly converges to the required dipolar soliton solution (also centered at the origin).  Note that a different choice for initial guess may lead instead to the homogeneous ground state.  

The absolute value of the phase is arbitrary, and to aid convergence we fix the value of the phase at one end of the grid.  The grid spacing $\Delta z$ is typically $0.1\xi$. Away from the phonon/roton instabilities, we employ a box of typical length $100\xi$.  However, close to these instabilities, box sizes of up to $1600\xi$ were required to ensure good approximation to the infinite limit (that is, for the mean-field dipolar potential to reach its homogeneous value at the boundaries).  The solutions were deemed converged when the relative residual, calculated as $||\bf{J}\delta\phi+\bf{f}||/||\bf{f}||$, had fallen below an arbitrary tolerance of $10^{-5}$.  

The above numerical method is akin to solving the equation $\bf{J}\delta\phi=-\bf{f}$ for $\delta\phi$, then setting $\phi^{(p+1)}=\phi^{(p)}+\delta\phi$ at each step $p$. The advantage of using the BCGM is that we only require knowledge of the transpose of the Jacobian, which is numerically faster to obtain than matrix inversion.

\end{document}